# The SSDC contribution to the improvement of knowledge by means of 3D data projections of minor bodies

Angelo Zinzi* [a, b], Mauro Ciarniello [c], Vincenzo Della Corte [d, c], Stavro Ivanovski [c, d], Andrea Longobardo [c,d], Alessandra Migliorini [c], Maria Teresa Capria [c], Ernesto Palomba [c], Alessandra Rotundi [d, c]

a) Space Science Data Center - ASI, Via del Politecnico snc, 00133, Rome, Italy

b) INAF-OAR, Via Frascati n. 33, 00078, Monte Porzio Catone (RM), Italy

c) INAF-IAPS, Via del Fosso del Cavaliere n. 100, 00133, Rome, Italy

d) Università degli Studi di Napoli "Parthenope", Dipartimento di Scienze e Tecnologie, CDN IC4, 80143, Naples, Italy

* Corresponding author: Angelo Zinzi. Address: Space Science Data Center - ASI, Via del Politecnico snc, 00133, Rome, Italy. Email: angelo.zinzi@ssdc.asi.it. Phone: +39 068567884

## Abstract

The latest developments of planetary exploration missions devoted to minor bodies required new solutions to correctly visualize and analyse data acquired over irregularly shaped bodies.
ASI Space Science Data Center (SSDC – ASI, formerly ASDC – ASI Science Data Center) worked on this task since early 2013, when started developing the web tool MATISSE (Multi-purpose Advanced Tool for the Instruments of the Solar System Exploration) mainly focused on the Rosetta/ESA space mission data.
In order to visualize very high-resolution shape models, MATISSE uses a Python module (vtpMaker), which can also be launched as a stand-alone command-line software.
MATISSE and vtpMaker are part of the SSDC contribution to the new challenges imposed by the "orbital exploration" of minor bodies: 1) MATISSE allows to search for specific observations inside datasets and then analyse them in parallel, providing high-level outputs; 2) the 3D capabilities of both tools are critical in inferring information otherwise difficult to retrieve for non-spherical targets and, as in the case for the GIADA instrument onboard Rosetta, to visualize data related to the coma.
New tasks and features adding valuable capabilities to the minor bodies SSDC tools are planned for the near future thanks to new collaborations.

## 1. Introduction

Recently a set of space missions for the first time observed several minor bodies (i.e., asteroids, comets and dwarf planets) for extended periods, mapping them at an unprecedented surface coverage and spatial resolution for objects included in this class.
Main examples of this new way of exploring minor bodies are ESA (European Space Agency) Rosetta, that orbited comet 67P Churyumov-Gerasimenko (hereafter called 67P, or "the comet"), and NASA (National Aeronautic and Space Administration) Dawn, that orbited two different bodies: asteroid 4 Vesta and dwarf planet 1 Ceres.

The scientific return of these missions is impressive, but researchers found new challenges to cope with. A major one was the necessity to map data belonging to non-spherical surfaces, in particular for comet 67P and, to some extent, asteroid 4 Vesta. For such irregularly shaped bodies, in order to easily retrieve good scientific information, data need to be visualized over their three-dimensional shape model, as classical bi-dimensional cartographic visualizations can be misleading.

Several groups dedicated their work to fulfil this task (e.g., Genot and Cecconi, 2014; Preusker et al., 2015; Klima et al., 2016).

Among them, the ASI Space Science Data Center (SSDC) established its Solar System Exploration (SSE) branch in late 2012 with the primary objective of providing a tool for visualizing data from VIRTIS-M, the Visible and Infrared Thermal Imaging Spectrometer (the final M stands for Mapper) onboard Rosetta (Coradini et al., 1997), by means of an interactive web-page displaying the shape model of the mission targets.

This tool, named MATISSE (Multi-purpose Advanced Tool for Instruments for the Solar System Exploration – Zinzi et al., 2016), presently allows queries to a georeferenced database comprising not only the already mentioned ESA Rosetta and NASA Dawn data, but also Chinese Chang'e missions to the Moon (Scaioni et al., 2016), NASA Messenger to Mercury, Mars Reconnaissance Orbiter to Mars and ESA Venus Express to Venus.

It is worthy to mention that all the observations accessed by MATISSE already underwent at least a 1B Data Processing Level (i.e., processed to sensor units) and are always accompanied by geometric information and metadata (e.g., longitude, latitude, phase/emergence/incident angles) computed by means of the appropriate NAIF (Navigation and Ancillary Information Facility) SPICE toolkit (Acton, 1996) by the scientific teams of the instrument. Therefore, the pipeline of MATISSE never computes geometries from SPICE and only in specific cases performs photometric corrections on data (e.g., Longobardo et al., this issue).

However, the SSDC-SSE effort did not end with the MATISSE tool development: major advances have been granted elaborating an "ad hoc" Python software (vtpMaker.py). This script, integrated into the pipeline, allows the 3D visualization of all the observations accessible by MATISSE (https://tools.asdc.asi.it/matisse.jsp). In addition it can be used as a stand-alone software for more complex data not present in the MATISSE database: examples can be multiple datasets merged together (Palomba et al., 2015), regularly gridded maps (Capaccioni et al., 2015), jets from the coma (Migliorini et al., 2016a) and inner coma dust properties (Della Corte et al., 2016).

This paper is organized as follows: Section 2 is dedicated to a brief description of the MATISSE tool and the vtpMaker Python module together with an overview of other similar tools; Section 3 shows the results obtained thanks to the 3D capabilities offered by the tool; in Section 4 conclusions and future developments are outlined.

## 2. Software infrastructure

### 2.1 MATISSE

MATISSE is the SSDC web-tool to access, analyse and visualize planetology data: since a dedicated paper has already been published (Zinzi et al., 2016), in this section we will only briefly describe its main features.

MATISSE mainly provides the possibility of searching a database by means of metadata, visualizing the results directly on the 3D shape model of the target.

It has been developed using different computer languages (e.g., bash shell, IDL/GDL, Python, C), looking for the best solution for every dataset could have been found. The database management system chosen is a MySQL one, made up of several tables, so that every useful information can be correctly managed and retrieved and an automated software can regularly synchronize the archive with the original external repositories. After the data selection has been defined by the user, a bash shell script generates the entire pipeline, choosing the scripts to be executed on the basis of the selected combination of instrument, observation and operation.

The main MATISSE advantages are:

- the structured database allows to rapidly search for specific observations in the dataset;
- the server-side pipeline makes it possible to transform the original observations stored in the database to both standard formats and high-level outputs, such as mosaics, ratios/differences (of different observations or of different spectral channels) and RGB false colour images;
- the user can hence download only the product required for its science objective;
- the user can extract information from the data almost ignoring the format in which they have been stored.

Even if some Planetary Data System (PDS – JPL, 2009) formatted observations could be opened directly with the same libraries as FITS (i.e., GDAL - Geospatial Data Abstraction Library), this is not a general case. GDAL has been developed mainly for imaging data, such as those acquired by visible cameras or imaging spectrometers and, "provides read/creation/update access to ISIS3 formatted imagery data"[1].

However the PDS format is designed to handle a very large variety of data types, not only those described by images. Furthermore, since up to PDS3 this standard specifically requires only that the way in which the file is recorded is declared in the header of the file (in human readable format), PDS files are not necessarily readable by a single software.

For example, at the present time no software based on GDAL is able to handle PDS files from Rosetta-GIADA (Grain Impact Analyser and Dust Accumulator), consisting of ASCII tables.

Therefore, for cases different from the imagery ones above mentioned, ad hoc software need to be developed in order to access a number of PDS formatted observations by means of GDAL libraries.

This could be critical for interoperability and data fusion issues, but by means of MATISSE pipeline, designed to read the different data types here discusses, it could be overcome, allowing the user to only search for the data to be inspected.

This capability could be used in the future to support upcoming missions, such as ESA BepiColombo, to be launched in 2018, thanks to direct collaborations with their scientific teams. In this way, the tool could be modified in order to improve scientific return of such missions.

MATISSE final output generally comprises a web page where a 3D visualization is associated to standard 2D static projections and two downloadable compressed files of the corresponding 2D and 3D offline visualizations (Fig. 1). This standard varies for different targets, e.g. for extended objects (Mars, Mercury, Moon and Ceres) no 3D online visualization is provided as it would result in a too much poor resolution.

Indeed, in order to preserve good network performances, the online outputs have lower resolutions compared to the offline ones and generally provide only a first guess of the operations applied by the user.

On the other hand, the files to be used offline offer extremely advanced scientific functionalities: they can be opened with standard 2D GIS (Geographic Information System) software (e.g., QGIS, ArcGIS,

---

[1] http://www.gdal.org/frmt_isis3.html

ENVI) or with 3D software using VTK (Visualization ToolKit) Python libraries (e.g., Paraview), thus giving the possibility for further enhancement of the data analysis already performed by MATISSE.

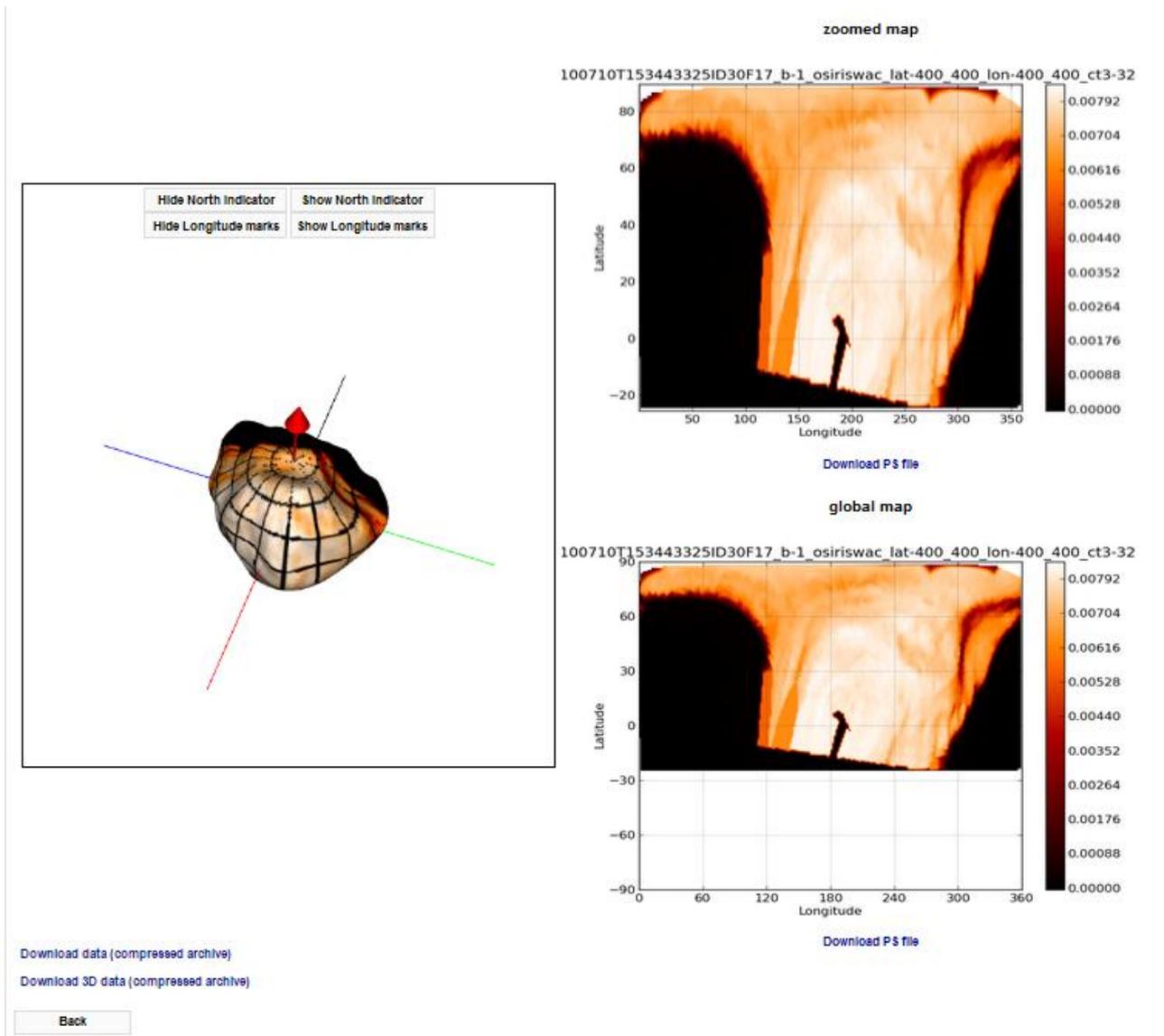

**Figure 1:** MATISSE output page for OSIRIS-Rosetta observations of asteroid 21 Lutetia.

## 2.2 vtpMaker

The core of the MATISSE 3D high-resolution visualization is the Python module vtpMaker[2] that, using as input the data ingested, read and formatted by MATISSE, provides as output a VTP (VisionTools Pro-E Project) file to be read with software using VTK Python libraries.
This script, originally developed for the VIRTIS-Rosetta data, now accepts different file formats and can be used either as part of MATISSE or as a standalone command-line software able to make 3D visualization of datasets not included in MATISSE.
The vtpMaker.py workflow can be summarized in five different steps:

1) reading file structures (i.e., matrix, vectors) from the input file;

---

[2] The vtpMaker.py code can be requested at the Solar System SSDC site http://solarsystem.ssdc.asi.it/vtpMaker.php, where also software requirements are listed.

2) selecting a geographical area of the observation to be further analysed;
3) importing the shape model of the target;
4) interpolating the datum from the original distribution to the shape model grid;
5) generating, by means of VTK Python libraries, the 3D VTP file to be read.

These tasks are performed using functions implemented inside the vtpMaker Python module (Fig. 2). The <main> function is the one dedicated to the acquisition of the input parameters and to the pipeline running, while the first function called by it is <dataReader>.

Here vtpMaker selects the appropriate reader for the different kind of files in input, thus storing the original observation values in the memory.

The pipeline subsequently comes back to the <main> function, starting the acquisition of the shape model / DTM (Digital Terrain Model) for the requested object, reading the values returned by <dataReader> (i.e., latitude, longitude and, in some cases, elevation of the observation) and then converting planetocentric (i.e., latitude, longitude, radius) coordinates to Cartesian (i.e., x, y, z) ones. With this input it is then possible to perform a Delaunay triangulation, computing surface polygons, using the <tridimsurf> function that makes use of scipy.spatial.Delauney Python function.

The next step is the spatial interpolation between the original observation and the shape model: this is done with one of the three functions <linear>, <nearest> or <cubic>, based on different cases, which make use of scipy.interpolate.griddata.

At this stage the data are ready to be written as output and, in order to save them in VTP format, the <makeVTP> function is called by vtpMaker. This function stores latitudes, longitudes and data values in the corresponding polygons of the shape model creating a vtkPolyData object and adding the required values as point data with the GetPointData function (latitude and longitudes are arrays, data values are scalars).

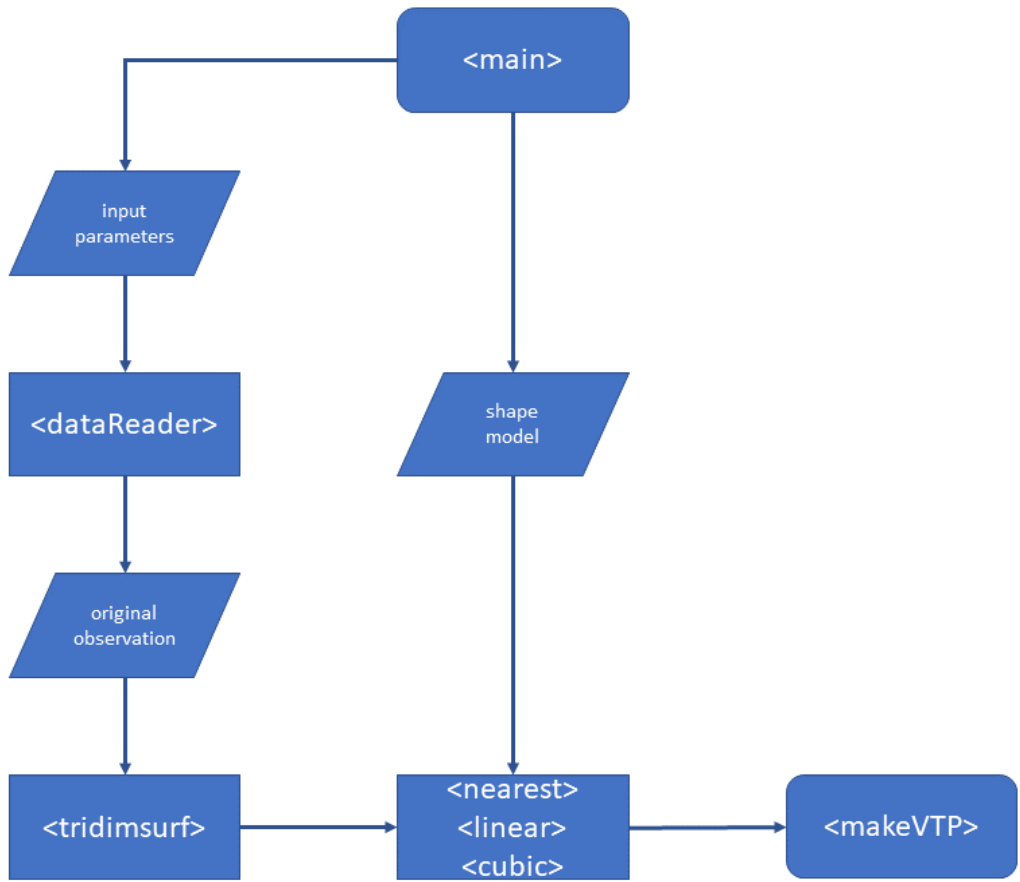

**Figure 2:** Flow chart of the vtpMaker module. The module functions are expressed as <function>

Required inputs are the observations to be projected and the shape model of the object, while the final output file contains three different fields (latitude, longitude and data values) for every shape model (or DTM) facet. At this stage of development the shape model is passed as a couple of binary files formatted (generally by conversion of wrl and obj common 3D formats) as in the following:

- planetocentric coordinates file (suffix "_plan.bin"): with two long integers describing respectively number of vertices and polygons of the shape model and two arrays of floats of length equal to the number of vertices for longitude and latitude, respectively;
- cartesian coordinates file (suffix "_cart-1.bin"): with the same two long integers for number of vertices and polygons as above, a float array of dimensions 3 x number of vertices for the vertices and an array of long integers with dimension 4 x number of polygons, whose fourth dimension is constantly set to -1 to indicate the end of the polygon.

These two file formats are preferred instead of the above mentioned common wrl/obj ones because the javascript used by MATISSE to visualize the shape model directly on the browser page requires a list of vertices and polygons exactly as stored in these two files.
Consequently, since vtpMaker is a software originally designed to be used inside the MATISSE pipeline, it presently requires them to compute the data interpolation. However an accompanying software is available on the SSDC web pages to transform wrl/obj files into the required formats.
Using the VTP file has many advantages with respect to both the low-resolution 3D online visualization and the high-resolution 2D offline visualization. In particular, the observation can be viewed directly on the very high-resolution shape of the target, thus allowing an easy way to grab information. In addition, the VTP format can be read by many different software using the VTK libraries, easily allowing further analysis and data processing, without be limited by commercial or software architecture issues.

Using the VTP files with software similar to Paraview, it is also possible to perform new kind of analysis, exploiting the software capabilities. For instance, Paraview has a large number of predefined statistical functions and allows generating custom filters by mean of Python scripting.

All these reasons, taken together, make the VTP format a very powerful way of storing large amount of data (even comprising those from an entire high-resolution shape model) in a way ready for advanced science analysis.

### 2.3 Comparison with other software with similar aims

Projection and visualization of imagery data in planetary science is a long-standing issue and therefore several tools have been already designed to cope with them. In particular, during latest years, they have been developed to be used on common personal computers, without extremely out of reach hardware requirements.

Among them the most widespread is probably the Integrated System for Imagers and Spectrometers (ISIS), developed since 1985 by the USGS (United States Geological Survey) for NASA and now at its version 3 (ISIS3). This is a free, specialized, digital image processing software package capable of placing many types of data in the correct cartographic location and enabling disparate data to be co-analysed.

It runs on many UNIX operating systems and can be used with a wide sample of space instruments and missions, from Apollo to the forthcoming OSIRIS-Rex (Origins Spectral Interpretation Resource Identification Security Regolith Explorer), while upgrades for upcoming missions are foreseen.

Another well-known software to manage imagery data is VICAR (Video Image Communication And Retrieval), whose history dates back to 1966, being still developed today by MIPL (Multimission Image Processing Lab) at NASA's Jet Propulsion Laboratory (JPL). As described in the "VICAR Quick-Start Guide" if you need to "do a variety of things that require maintaining precise scientific calibration of the data, VICAR is a better bet"[3] compared to Photoshop.

Similar to ISIS3 this software is intended to be used mainly with cameras and with some imaging spectrometer (e.g., CASSINI-VIMS) and, among its main advantages, are the possibility of reading directly PDS files and files bigger than 2 GB.

It officially runs on Linux and Solaris 10 machines, but is also supported by Mac OS X.

However these two tools, albeit of great relevance for the treatment of images and designed to manage projections of irregular bodies, are constrained to the classical bidimensional visualizations, thus limiting their usefulness for the irregular shapes of the great part of minor bodies. Therefore, other solutions with native 3D packages have been recently proposed specifically for solar system objects by other organizations, such as J-Asteroid (Smith et al., 2015 –derived from JMARS) and the Small Body Mapping Tool (SBMT, Klima et al., 2016).

SBMT seems to have more than one point of contact with the work here described, having been designed specifically for small and irregular bodies and using Paraview (Ahrens et al., 2004) to show the target object, but it cannot import datasets not already present in the downloaded package, thus limiting its capabilities.

---

[3] https://www-mipl.jpl.nasa.gov/vicar_os/v1.0/vicar-docs/VICAR_guide_1.0.pdf

This is the most limiting difference between this tool and that developed by SSDC: vtpMaker can indeed process any observation provided in one of the accepted formats, such as ENVI, a standard (composed of a couple of files with data and header) largely used by the terrestrial and planetological GIS community.

The last tool here reviewed (J-Asteroid) is composed of a jar executable file and this makes it multiplatform running on Windows, MacOS and Linux. It is derived from JMARS, a GIS developed by Arizona State University's Mars Space Flight Facility. In literature it is described as being adapted for only two missions, nominally Dawn and OSIRIS-REx, and for these missions it can be used for both visualizing data and planning future observations.

One of the main advantages of using J-Asteroid is the possibility of looking for JPL datasets and visualize them directly on the shape model (stored in the BDS NAIF SPICE standard).

This makes it similar to MATISSE as a data discovery tool able to search for data in archives and visualize them in both 2D and 3D. However, one of the main differences between it and MATISSE is that the SSDC tool is also a web-based data visualization tool, not requiring any installation nor downloading of initial data: only the final output files are saved on the local machine by the user, when needed, after the visualization of the result on the web page.

## 3. Science advances allowed by the use of 3D visualizations

For spherical or quasi-spherical bodies classical bi-dimensional data visualizations are often sufficient to allow a comprehensive analysis in both space and time.

Minor bodies, on the contrary, share some characteristics that make 2D visualization not always accurate enough: surface data generally have irregular shapes, whereas observations acquired around the objects have distances from the surface comparable to its size (differently from the case of planets, where the atmosphere is several orders of magnitude smaller than the radius of the planet).

Therefore, in this case, a three-dimensional visualization is required to correctly understand the studied environment and to gather useful pieces of information, as clearly demonstrated recently by several scientific results from missions devoted to the study of minor bodies.

Some of them are pointed out in the following examples, where, without the support of a correct 3D visualization, the analysis could have been partial or even misleading.

### 3.1 Comet 67P nucleus

As soon as the visible cameras onboard ESA Rosetta spacecraft showed the extremely irregular shape of comet 67P nucleus (Sierks et al., 2015), the need of using a different approach with respect to a bi-dimensional data analysis and visualization clearly emerged.

A 3D data visualization provided a better characterization of the 67P bi-lobate nucleus, as it made possible to correctly estimate the exact position of specific regions on the surface and their real extension.

One of the best examples concerns the Imothep region on the large lobe and the Hatmehit one on the small lobe (El-Maarry et al., 2016), which are not easily recognized as distant in space as they really are using a classical 2D visualization. In addition, the use of a 3D representation allows to visualize areas that would be hidden in the 2D maps. For instance, the Hator region, located on the small lobe in proximity of 67P's neck (the portion connecting the two main lobes), is not fully visible in the classical bi-dimensional maps because of the concave shape of the nucleus (Fig. 3).

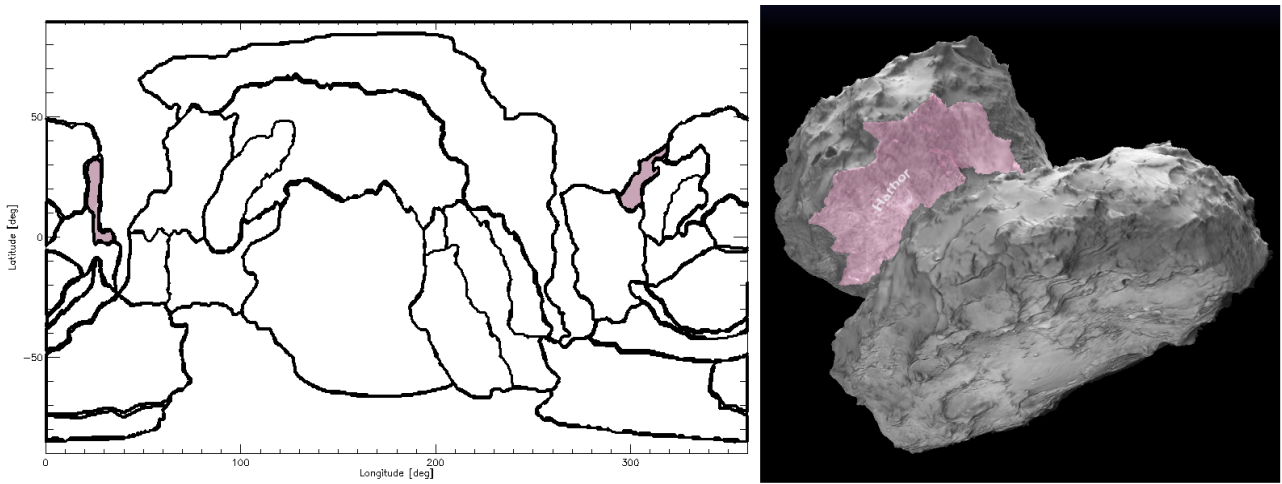

**Figure 3:** Comparison between 2D (left – adapted from El-Maarry et al., 2016) and 3D (right – obtained by http://sci.esa.int/comet-viewer/) projections of the morphological regions of 67P: the Hator region (pink coloured in both images), largely hidden by the small lobe in the 2D projection, is clearly recognizable in the 3D image.

The SSDC expertise in 3D visualization was applied to the very first VIRTIS-Rosetta data, e.g. the 67P surface spectral slope at visible and infrared wavelengths was mapped onto the nucleus shape model, providing a first indication of a larger content of water ice in the "comet neck" (Figure 2 in Capaccioni et al. 2015). This is compatible with the results reported in Quirico et al. (2016) where a 3D map of the intensity of the ubiquitous 3.2 µm absorption feature, related to organic compounds, is shown for different observations, indicating an increase of the band depth in the neck. The deepening of the absorption can be indeed explained by a larger amount of water ice, which contributes to the intensity of the feature, mixed with the surface material

Along with the water ice spatial distribution, the use of 3D re-projection techniques also improved the characterization of its temporal variability and its dependence on the illumination conditions, as reported in the movies published on the ESA Rosetta blog (http://blogs.esa.int/rosetta/2015/09/29/an-update-on-comet-67pc-gs-water-ice-cycle/, Ciarniello et al. 2015 – Fig. 4), showing the surface water ice diurnal cycle inferred from VIRTIS data (De Sanctis et al., 2015, Ciarniello et al., 2016). Furthermore, in Ciarniello et al. (2016), a 3D mapping of regions of interest helped for the identification of their position on the nucleus and their real size.

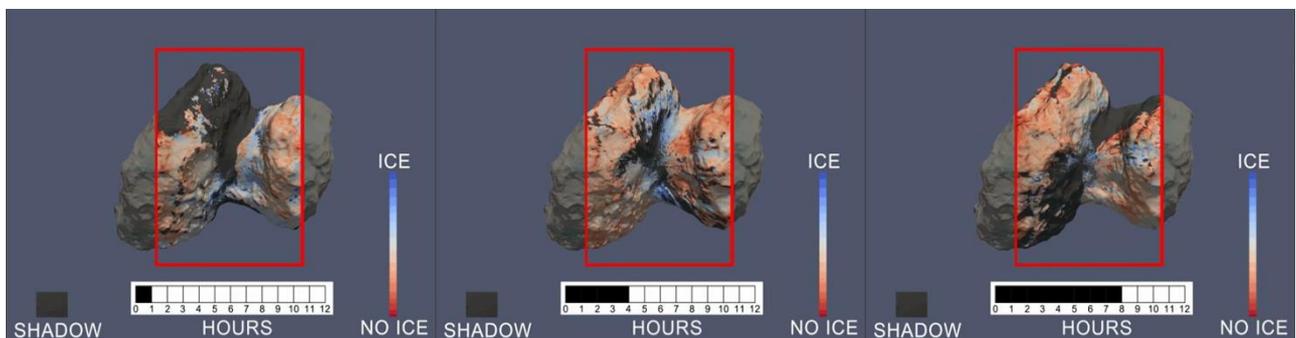

**Figure 4:** Three different snapshots of the video presented by Ciarniello et al., 2015 (http://blogs.esa.int/rosetta/2015/09/29/an-update-on-comet-67pc-gs-water-ice-cycle/) showing the water ice diurnal cycle on the 67P surface.

### 3.2 Gas and dust particles ejections from 67P nucleus

One of the main topics for cometary sciences deals with the investigation of the coma dust-gas environment.

In this case 2D maps, even using different colours, opacities, line weights and symbols, hardly show in a straightforward way the peculiarities of this environment. On the contrary, 3D visualizations can be of valuable support, as they are able to accurately display the shape and spatial distribution required for a detailed analysis.

A good example is the representation of the water and carbon dioxide emission distributions around the 67P nucleus, as reported in Migliorini et al. (2016a). In that paper only 2D projections are used, either showing data as observed in the VIRTIS imaging spectrometer Field Of View, with the comet superimposed (Fig. 5a), or as map projections, with coloured stripes indicating the observed areas and the abundance of the studied compound.

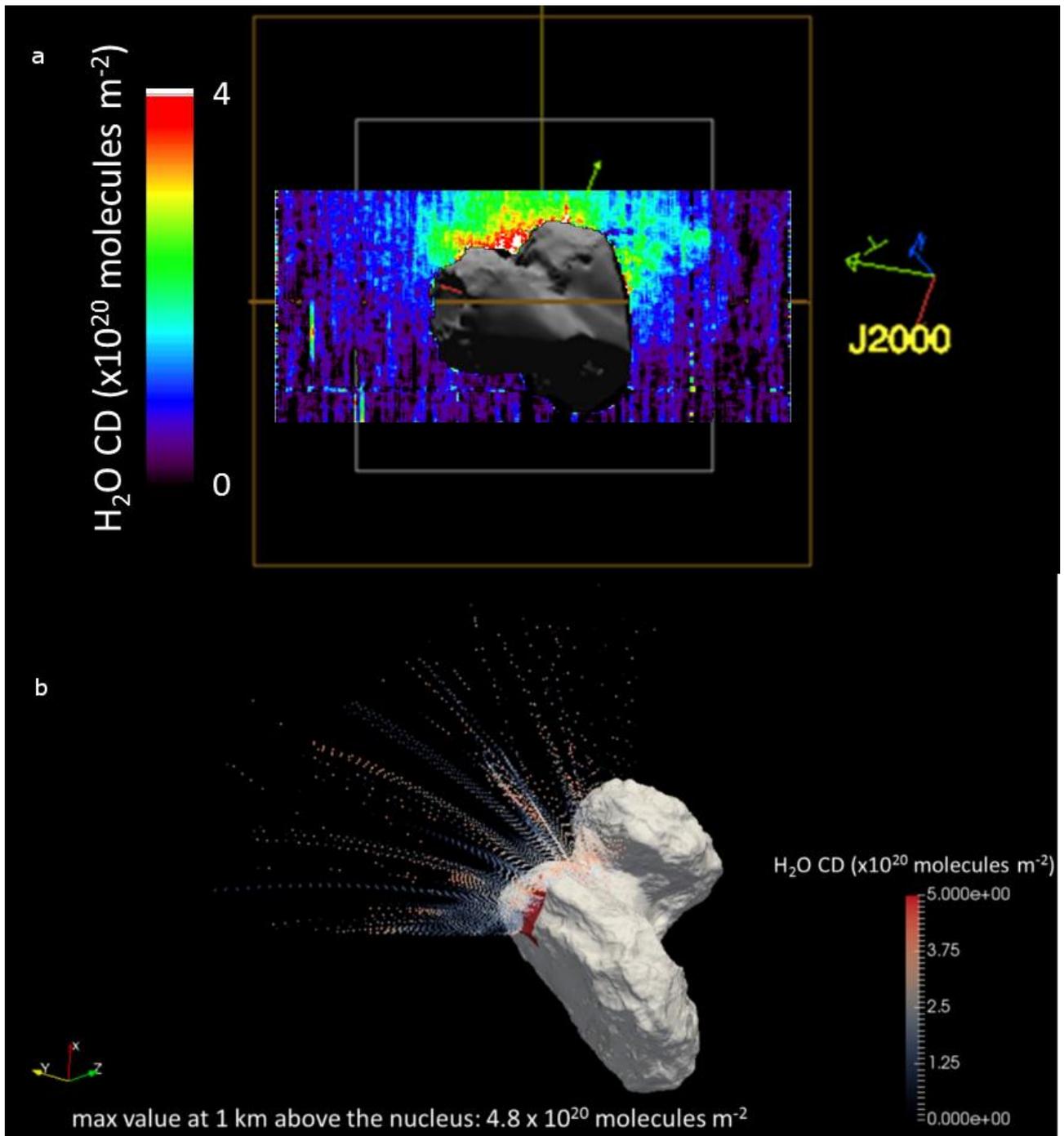

**Figure 5:** Comparison between of a single VIRTIS-Rosetta observation of water column density for: a) 2D projection; b) 3D projection.

The use of a 3D visualization (Migliorini et al., 2016b) permitted inferring additional information from the single VIRTIS observations (Fig. 5b), adding the possibility of locating the region where emission is originated. This became possible by using a "pseudo-triangulation" (i.e., the use of different observations belonging to the same phenomenon and area at different times): in particular the 3D projection of three different VIRTIS observations, acquired after about 1 comet rotation, made it possible to identify a region which is likely to be the source area of the emitted gas. As it is evident by looking at Fig. 6, with this visualization the gas emission seems to really depart from the comet nucleus depicting a cone-shaped volume near it, perfectly giving the idea of a cometary jet.

Even though this exercise cannot be conclusive, because a real triangulation of images would be necessary to link gas to active regions (Vincent et al., 2016), it is out of doubt that, without this 3D visualization, constraining the nucleus regions linked to the observed $H_2O$ and $CO_2$ emissions would have been extremely difficult.

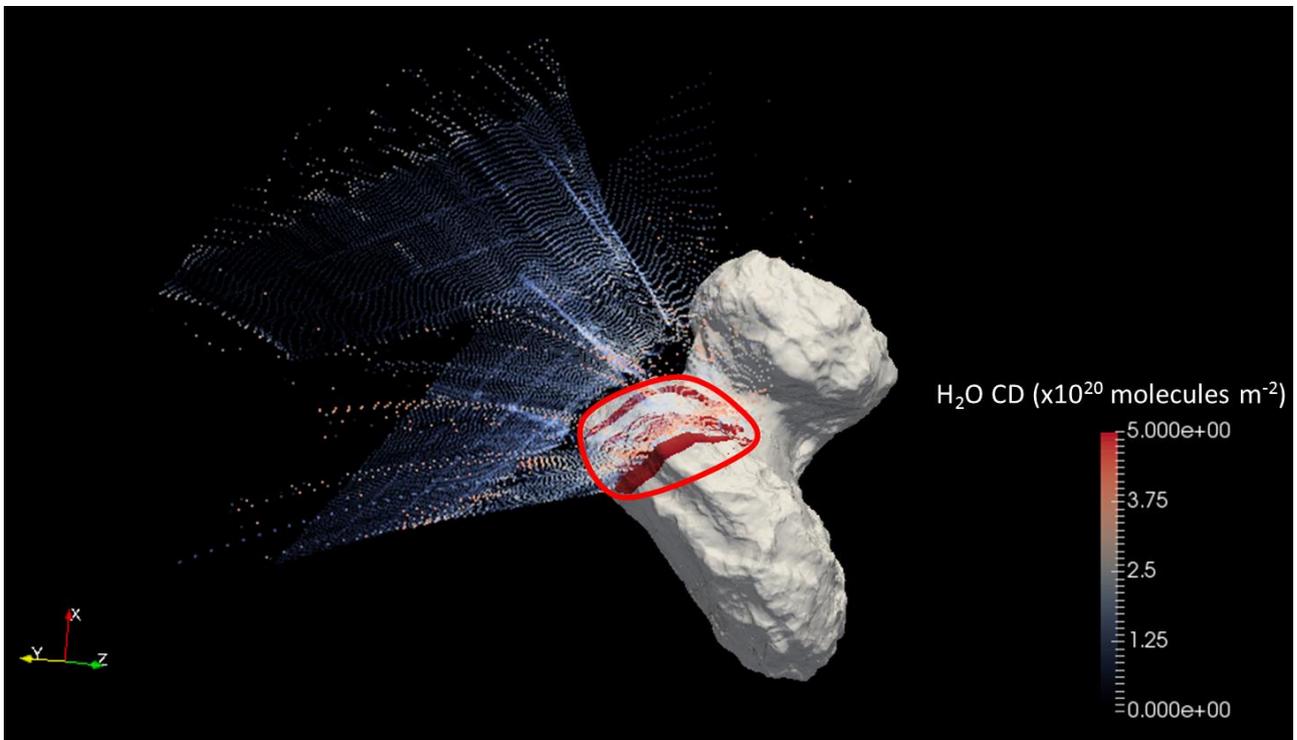

**Figure 6:** Water column density obtained for images I1_00387385977, I1_00387426413, I1_00387561494, acquired consecutively and spanned with a finite number of comet's rotations (the second image is acquired after about 1 comet's rotation, and the last one after about four comet's rotations with respect to the very first one). The three images observe approximately the same region, allowing identifying an area where the emission seems to originate from.

Another Rosetta instrument that benefits of a 3D visualization for the environment encircling the comet is GIADA (Colangeli et al., 2007; Della Corte et al., 2014; 2016a), designed to characterise the dynamical properties of individual dust particles in the 67P coma (Rotundi et al., 2015).
Along with the dust particle speed, optical cross section and momentum, GIADA measurements provide the detection time (in UTC) and the position of the detected particle in the coma. The 3D visualization of the detected particles along the Rosetta's trajectory allows describing the dust distribution in the coma.
In addition, a 3D plot can be compared to the results of a 3D numerical simulations obtained with dust dynamical models. Della Corte et al. (2016) clearly showed that a proper data visualization, e.g. with the appropriate choice of the reference coordinate system, is critical to reconstruct the coma dust spatial density.
As an example of 3D GIADA data visualization, we report in Fig. 7 the histogram of particle detections along the Rosetta trajectory occurred on February 22$^{nd}$ and 23$^{rd}$, 2015. In order to demonstrate the valuable contribution that 3D visualization could provide for the GIADA science branch here we show 3D visualisation of all GIADA detections grouped at regular time intervals (around 5 seconds) along the Rosetta trajectory for two dates with many detections in order to present the detections of the dust particles: February 22nd and 23d, 2015 (Fig. 6).

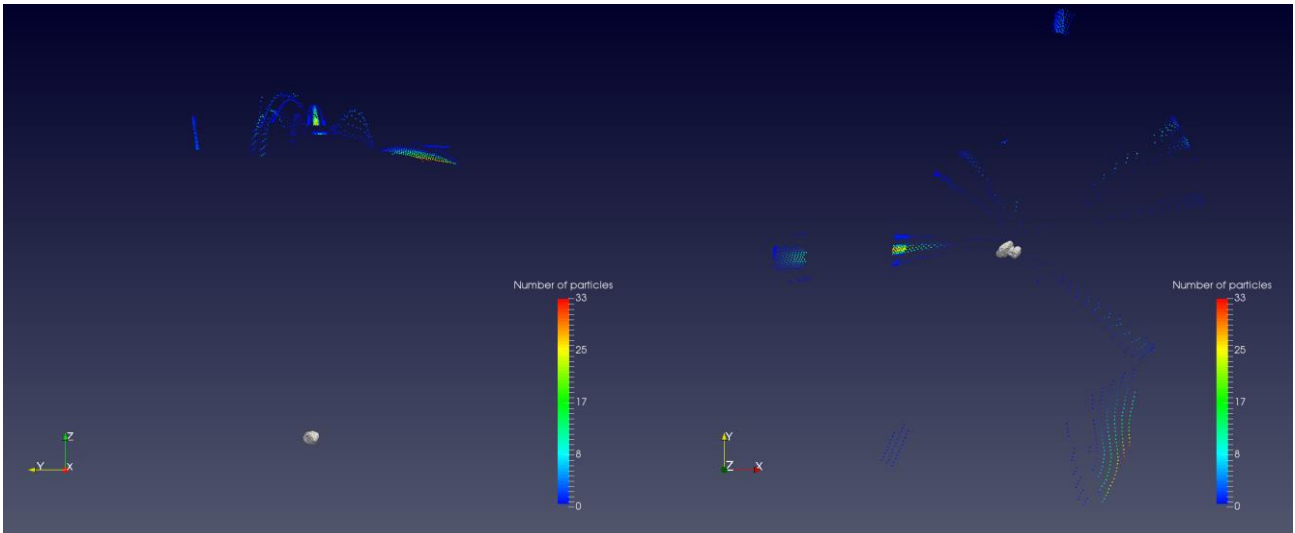

**Figure 7:** Two different views of the data acquired by GIADA on 22nd and 23rd February 2015, where the exact location of the particles in the cometocentric reference system is shown. We used a time step of acquisition of 5 minutes.

### 3.3 Data fusion

Even if the targets of the NASA Dawn mission are not as small and irregular as the 67P, applying the 3D visualization software developed by SSDC to data already georeferenced resulted to be of great importance. In particular, it was critical to fuse different datasets improving in number and accuracy the information on surface mineralogy of asteroids 4 Vesta and 21 Lutetia (e.g., Palomba et al., 2015; Longobardo et al., 2015).
In the first case Digital Elevation Models (DEMs), albedo maps retrieved by Dawn Framing Camera and infrared band depths from VIR (Visible and InfraRed spectrometer) imaging spectrometer were merged together. Albeit the visualization of the albedo and visible/infrared spectral parameters directly on the topography cannot alone indicate the systematic presence of the studied minerals, it improved the possibility of contextualizing the search for minerals generally associated to specific topographic features (e.g., crater rims) (Fig. 8) or surface colour variations (i.e., dark or bright areas).

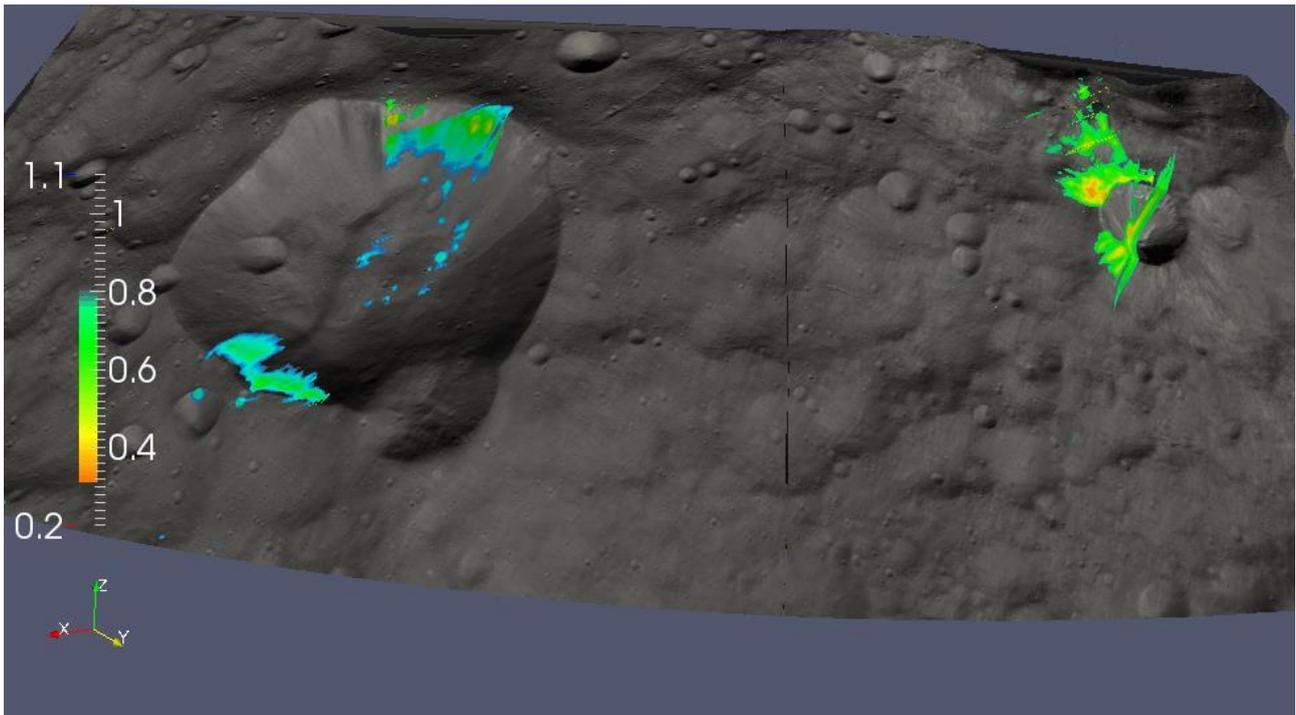

**Figure 8:** Band Area Ratio (BAR) visualization for the Arruncia-Bellicia field on Vesta. This spectral parameter has been used by Palomba et al., (2015) in the identification of olivines on Vesta.

Another important application of 3D visualization has been with Rosetta-VIRTIS data acquired during the asteroid Lutetia fly-by performed on the way to 67P (Longobardo et al., 2015). The 3D distribution of geometric albedo on the Lutetia's shape model allowed recognizing tectonic linear features (i.e., grooves) associated to albedo markings (Fig. 9), an uncommon behavior, which has been ascribed to subsurface exposition or downslope movements.
In particular, the use of VTP outputs in Paraview allowed the combination of different colouring for the different datasets (e.g., grayscale for albedo and coloured for IR band depths), thus improving the localization of narrow deposits, for example by using ad-hoc scales and transparencies.

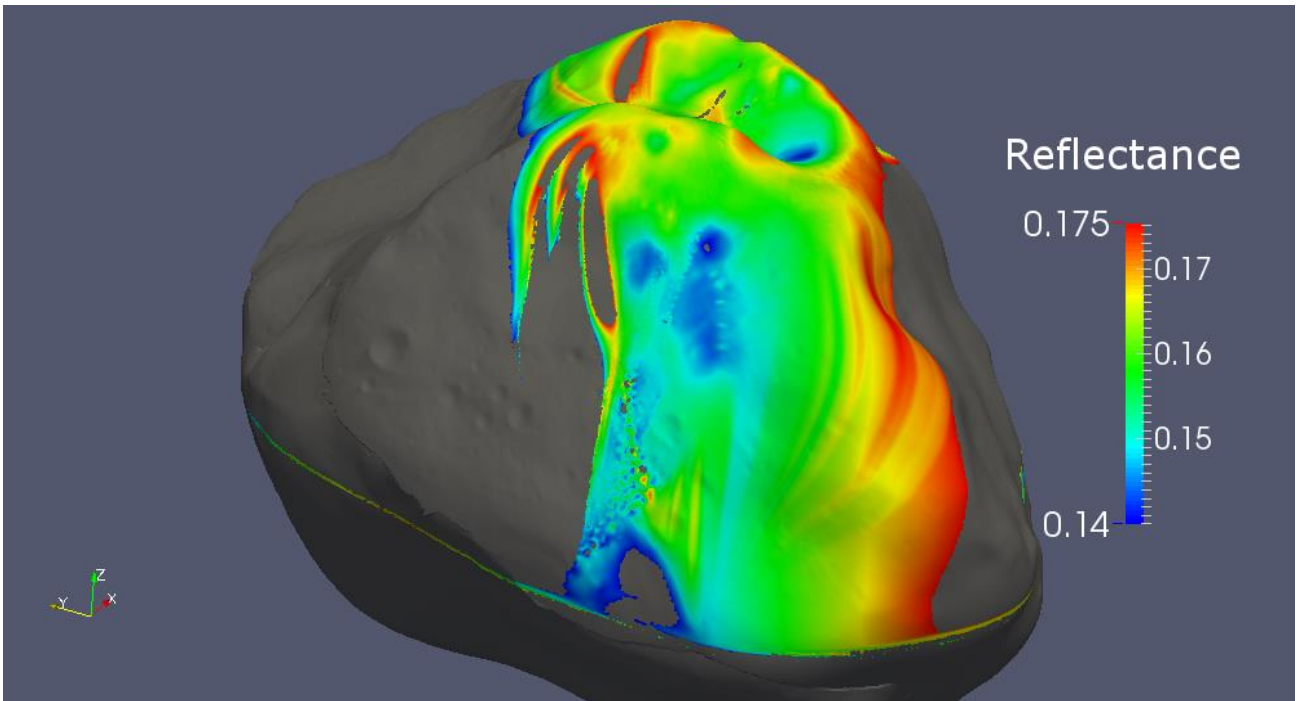

**Figure 9:** 3D visualization of albedo differences over asteroid 21 Lutetia, linked to surface features, similar to those published by Longobardo et al. (2015).

### 3.4 Outreach

The ease in using MATISSE has been demonstrated during the event organised at the Italian Space Agency HQ in June 2016 in the frame of "Alternanza Scuola-Lavoro" (i.e., "School and Work synergy"), a project developed by the Italian Ministry for Education, University and Research (MIUR).
In this context, a group of high-school students actively used the tool to investigate data acquired by the VIR-Dawn instrument around Vesta (public dataset). Their assignment was to characterize the dark units present on Vesta surface mimicking the work performed by Palomba et al., (2014) (http://solarsystem.ssdc.asi.it/news.php).
In order to fulfil this task they used MATISSE to search data from specific regions and to evaluate the 1.98 μm VIR radiance. The selected data were then inspected in 3D and 2D, using the Paraview and JS9 software, respectively.
The activity performed by the students showed that also non-professional users can easily obtain reliable results in short time thanks to MATISSE.

### 4. Conclusions and future developments

In the context of the present day Solar System exploration, where minor bodies acquired a major portion of the funded missions, the use of the multi-platform 3D visualization suit of software developed by SSDC has been demonstrated to provide valuable consequences on the related science. This is particularly true for small and irregular objects, such as the comet 67P Churyumov-Gerasimenko: the data acquired by VIRTIS and GIADA instruments on-board the ESA Rosetta mission would have been hardly interpreted with a classical 2D projection or a spherical one. The projection of the data directly on the three-dimensional shape model of the 67P sped up their scientific interpretation and contributed to reach the astonishing Rosetta scientific objectives.

For larger and more regular objects, the possibility of visualizing spectral parameters together with topography and albedo variations generally associated to the minerals object of the studies, helped in better localize and characterize specific mineral deposits.

In all these cases the possibility of easily extracting scientific information from data stored in organized databases resulted to be of extreme importance in the context of a "data fusion perspective", where multi-instruments data have to be combined to increase the scientific return.

The forthcoming tasks to improve the SSDC MATISSE tool are related to new scientific and technological challenges, for example by:

- integrating the high-resolution and detailed analysis functionalities in the online version of MATISSE;
- using a different Database Management System (DBMS) to better exploit the GIS potentiality of the data;
- visualizing the 2D online maps directly using interactive software integrated in the MATISSE's output page, such as JS9, thus allowing an "a posteriori" customization of the visualization also for this type of result.

These tasks will be pursued also intensifying collaborations, so that it will be possible to improve the characteristics of our software, making it more useful for scientific advances in the study of minor bodies of the Solar System.

Examples could be those with scientific teams involved in ongoing and future missions directed to minor bodies (Longobardo et al., this issue,; Longobardo et al., 2017) and active participation to international programs aimed at increasing the use of open data and tools in order to facilitate the extraction and preservation of science information present in the datasets (e.g., Planetary VO – Erard et al., 2018 – Open Planetary – Manaud et al., 2016 – and Open Universe – Giommi et al., in press).